# Multimodal Emotion Recognition by Fusing Video Semantic in MOOC Learning Scenarios

Yuan Zhang [c], Xiaomei Tao [a, b, c, *], Hanxu Ai [c], Tao Chen [d], Yanling Gan [a, b, c]

[a] Key Lab of Education Blockchain and Intelligent Technology, Ministry of Education
[b] Guangxi Key Lab of Multi-Source Information Mining and Security
[c] School of Computer Science and Engineering, Guangxi Normal University
[d] School of Computer Science, University of Birmingham

**ABSTRACT**

In the Massive Open Online Courses (MOOC) learning scenario, the semantic information of instructional videos has a crucial impact on learners' emotional state. Learners mainly acquire knowledge by watching instructional videos, and the semantic information in the videos directly affects learners' emotional states. However, few studies have paid attention to the potential influence of the semantic information of instructional videos on learners' emotional states. To deeply explore the impact of video semantic information on learners' emotions, this paper innovatively proposes a multimodal emotion recognition method by fusing video semantic information and physiological signals. We generate video descriptions through a pre-trained large language model (LLM) to obtain high-level semantic information about instructional videos. Using the cross-attention mechanism for modal interaction, the semantic information is fused with the eye movement and PhotoPlethysmoGraphy (PPG) signals to obtain the features containing the critical information of the three modes. The accurate recognition of learners' emotional states is realized through the emotion classifier. The experimental results show that our method has significantly improved emotion recognition performance, providing a new perspective and efficient method for emotion recognition research in MOOC learning scenarios. The method proposed in this paper not only contributes to a deeper understanding of the impact of instructional videos on learners' emotional states but also provides a beneficial reference for future research on emotion recognition in MOOC learning scenarios.

## 1 Introduction

In recent years, emotion recognition in MOOC learning has received much attention. Although MOOC learning has advantages such as transcending time and space constraints and abundant learning resources, the problem of high dropout rates remains prominent [2, 18], and emotional deficiency is one of the essential reasons [26]. Emotions play a regulatory and mediating role in cognitive processes [29]; therefore, in the fields of education and cognitive science, understanding and managing emotions is crucial for optimizing learning strategies [9] . Research has shown that the presentation of different content in videos can affect the activity of the emotion regulation areas in the human brain, thereby influencing an individual's emotional state [14, 32] . Some studies incorporated physical level information such as brightness and saturation from videos into emotion recognition tasks [37] , which effectively improve accuracy, but fail to pay attention to the higher-level semantic information related to content carried by videos. At present, most research on affective recognition in MOOC learning often overlooks the potential impact of semantic information in instructional videos on learners' emotions.

Early research focused on analyzing students' engagement in learning through classroom tests and student evaluations [12, 20]. More and more studies have been conducted to identify learners' emotional states by acquiring signals such as facial expressions, eye movements, PPG, Electroencephalogram (EEG), Electrodermal Activity (EDA), etc. during the learning process [3, 23, 36, 37] . And achieve accurate identification of learners' emotional states by fusing information from multiple modal data at the data level, feature level, and decision level [3, 4, 21, 28]. In MOOC learning scenarios, learners' emotions are closely related to learners' personal cognition and semantic information in videos, and the presentation of different contents in teaching videos will bring different feelings to learners [11, 19, 22]. Therefore, we hypothesize that in MOOC learning scenarios, learners' emotions are closely related to the semantic information of instructional videos.

We propose a multimodal emotion recognition method based on the above analysis by fusing video semantic information and physiological signals. Specifically, we extract semantic information from instructional videos and fuse it with eye movement and PPG signals through the cross-attention mechanism to improve the performance of emotion recognition in MOOC learning. To the best of our knowledge, we are the first to apply the semantic information of instructional videos to emotion recognition tasks in MOOC learning, providing a new perspective for emotion recognition research in MOOC learning scenarios. The main contributions of this article can be summarized as follows:

- We hypothesize there is a close correlation between the semantic information of instructional videos and learners' emotions. To deeply explore the impact of video semantic information on learners' emotions, this study innovatively proposes a multimodal emotion recognition method that integrates video semantic information and physiological signals. The instructional video's high-level semantic information is obtained by generating video descriptions, which are fused with eye movement and PPG signals to identify the learner's emotional state. This method effectively improves the performance of emotion recognition.
- To effectively capture the correlation and complementarity between different modal features，we propose a multi-modal emotion recognition module based on cross-attention fusion. By first learning the feature representations of any two modalities separately and then further fusing the learned features to obtain the feature representations of three modalities, we achieved an effective fusion of the three modalities.
- We conducted extensive experiments and analyzed the experimental results in depth. The effectiveness and feasibility of our method have been comprehensively verified, and experimental results show that our method has achieved significant results in practice, with an accuracy improvement of over 14%.

* Corresponding Author. Email: xiaomei.tao@gxnu.edu.cn

The rest of this paper is organized as follows: Section 2 reviews previous work on emotion recognition. Section 3 provides a detailed explanation of the proposed method framework and the extraction of video semantic information. Section 4 reports and analyzes the experimental results, extensively verifying the effectiveness of the method proposed in this paper. Section 5 summarizes the experimental results and future work.

## 2 Related Works

### 2.1 Emotion Recognition Using Contextual or Semantic Information

Adding context or semantic information to emotion recognition tasks has gradually become a focus of attention. After adding context information, emotion recognition systems can more accurately infer related emotions. Kosti et al. [15] believe that in addition to facial expressions and body postures, scene context also provides important information for us to perceive people's emotions. Scene context information is also a key component in understanding emotional states. Therefore, they created and released the Emotions In Context (EMOTIC) dataset and proposed a baseline CNN model for emotion recognition in scene context. Dashtipour et al. [7] proposed a context-aware multi-modal sentiment analysis framework to predict emotional states accurately.

In addition, some studies use speech transcription to text to obtain additional contextual or semantic information based on audio and visual information. Jiang et al.[13] proposed a fuzzy temporal convolutional network based on context self-attention (CSAT-FTCN) to improve the effect of emotion recognition by using speech-transcribed text as a new modality and integrates it with the original audio and visual modalities. Xia et al. [35] transcribed speech into the text as semantic information to enhance audio and visual features. Meanwhile, semantic information also serves as a new modality for decision fusion with audio and video modalities for emotion recognition. Tzirakis et al. [30] enhanced the performance and effectiveness of emotion recognition by transcribing speech into text as semantic information in speech emotion recognition tasks. The above research shows that incorporating contextual or semantic information into emotion recognition tasks has a positive effect. We believe that further incorporating the high-level semantic information of videos as a global factor into MOOC learning scenarios will also have a positive effect on learners' emotion recognition task.

### 2.2 Multimodal Emotion Recognition Based on Attention Mechanism

In the field of emotion recognition, multimodal fusion is a challenging task. Early research typically used traditional data level, feature level, or decision level fusion methods [16, 33]. However, with the rise of attention mechanisms, research focus has gradually shifted towards cross-modal interaction [27, 35]. For example, Wang et al. [34] utilized an attention-based fusion emotion transformer fusion (ETF) framework to integrate features from EEG and eye movement signals. Xia et al. [35] designed a semantic enhancement module based on the attention mechanism, which enhances audio and visual features through semantic information. At the same time, semantic information is also integrated with audio and video as a new modality to improve emotion recognition performance. Gong et al. [10] proposed an intra- and inter-modality attention fusion network that effectively learns the critical information between the two modalities and improves the effectiveness of emotion recognition. These studies show that using attention mechanisms can better learn the correlation and complementarity between different modalities, thus achieving more effective multimodal fusion effects.

Based on the semantic information in instructional videos, we propose a multimodal multimodal emotion recognition method by fusing video semantics and physiological signals. By generating video descriptions, we obtain the high-level semantic representation of instructional videos and use cross-attention mechanisms to fuse them with eye movement and PPG signals, effectively improving the performance of MOOC learning emotion recognition.

## 3 Proposed Method

### 3.1 Overall Framework of The Method

The multimodal emotion recognition method by fusing video semantic information and physiological signals consists of three main stages: data processing and feature extraction, cross-attention fusion, and emotion classification (as shown in Figure 1). Firstly, in the data processing and feature extraction stages, we preprocess and extract physiological signals and video semantic information respectively. The physiological signals take eye movement and PPG signals as examples, while video semantic information is derived from the extraction of video stimulus materials in video learning scenarios. Then, the extracted eye movement, PPG, and video semantic features are fed into the fusion module. The fusion module is mainly based on the cross-attention mechanism, which combines the features of three modalities in any pairwise manner and inputs them into multi-head attention to learn the corresponding feature representations. Then, these features are further fused to obtain features that contain important information for three modalities. Finally, the fused features are input into the sentiment classifier for final sentiment prediction.

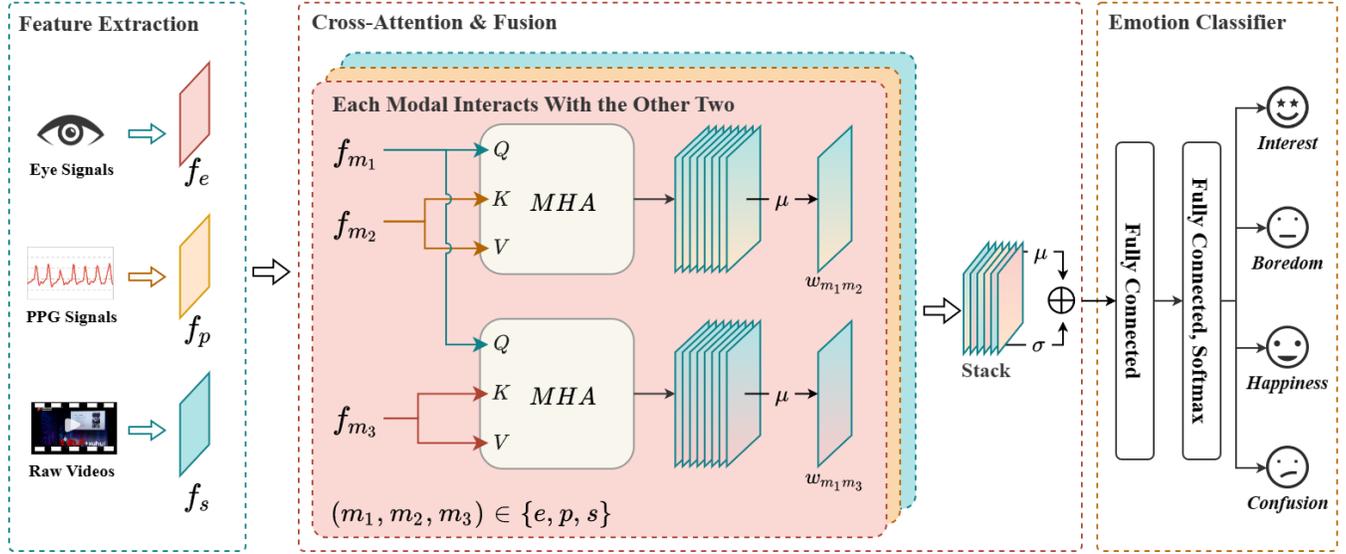

Figure 1. It consists of three main stages: data processing and feature extraction, cross-attention fusion, and emotion classification. Among them, $f_e, f_p, f_s$ represent eye movements, PPG and semantic features respectively, $m_1, m_2, m_3$ represent the three modalities, the symbol $\mu$ represents the mean, and $\sigma$ represents the standard deviation

### 3.2 Video Semantic Information Generation and Feature Extraction

In this study, a crucial task is to obtain semantic information from videos and extract features from video semantic information. To achieve this goal, we first use pre-trained LLM to generate video descriptions to obtain semantic information in instructional videos. Then, we used the pre-trained BERT model [8] to extract the features of video semantic information. The specific process is shown in Figure 2: Firstly, to ensure the stability of the subsequent running process, we preprocess the original videos and convert them to a unified resolution (1280 x 720), frame rate (25fps), and target bit rate (1000k). Then, we feed the videos into the pre-trained model mPLUG-Owl [38] (The mPLUG-Owl model is available on the GitHub, HuggingFace, or ModelScope platforms) trained on LLM for generating video descriptions to obtain semantic information. The automatic generation of semantic information in instructional videos has been achieved through the mPLUG-Owl model. As shown in the example in Figure 2, the generated semantic information includes key content such as scenes, objects, actions, and plots in the video.

For extracting video semantic features, we first perform text cleaning on the obtained semantic information and then put the preprocessed video semantic information into the Bert model for feature extraction. Semantic information is encoded in tokenized form during this process, and positional encoding is added to each token. Subsequently, after processing by multiple Transformer encoder layers, the model utilizes the self-attention mechanism and feedforward neural network to capture the semantic relationships between tokens. Next, feature representations are extracted from the last Transformer encoder layer to obtain high-quality semantic features. Due to the high dimensionality of the features extracted using the Bert model, we applied the PCA algorithm to reduce the dimensionality of semantic features. The feature dimensions were reduced to 20, 25, 50, 70, and 100, respectively, and experiments showed that the best effect was achieved at 25 dimensions. Therefore, we chose to reduce the semantic features to 25 dimensions and further use LSTM for encoding to obtain the final video semantic features for subsequent experiments.

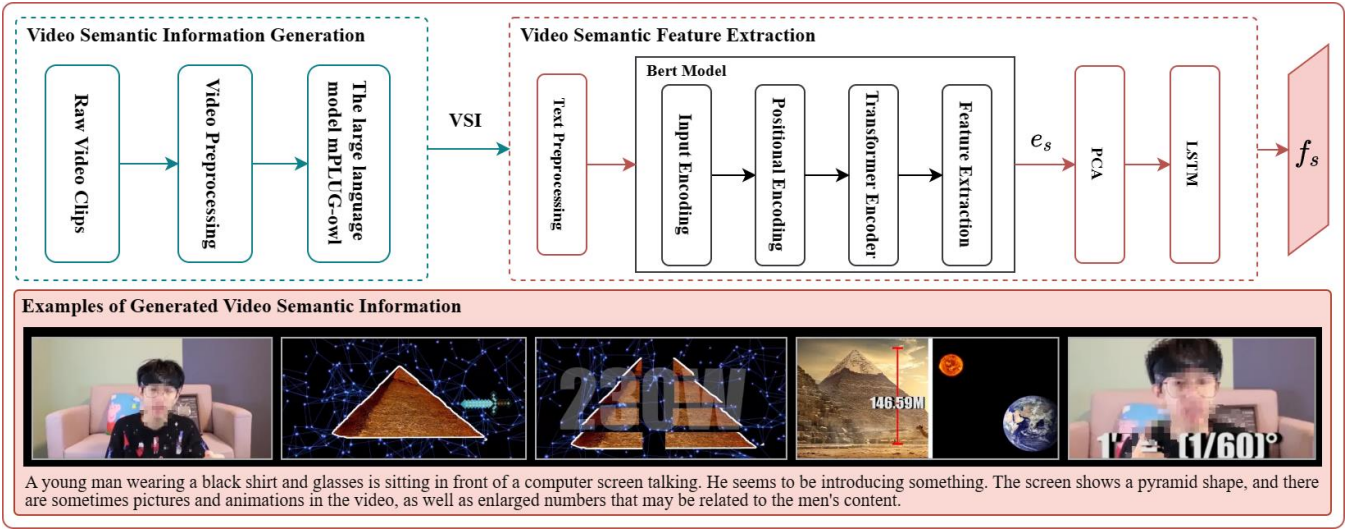

Figure 2. Schematic diagram of video semantic information generation and semantic feature extraction

## 3.4 Cross-Attention Fusion

To effectively learn important information between different modalities, we use Multi-Head Attention (MHA) to model the cross-attention fusion module and use MHA to learn information between any two modalities separately. Each MHA module requires three inputs, namely Query(Q), Key(K), and Value（V）. In this article, when learning information from two modalities, we use one modality as the input for Q, while the other modality serves as both K and V. Each input is first projected into a different subspace using a linear layer H times, where H represents the number of heads. The projection of each subspace $h \in \{0, \ldots, H-1\}$ is expressed as:

$$Q_h = W_h^Q e_{m_1}, \quad (1)$$
$$K_h = W_h^K e_{m_2}, \quad (2)$$
$$V_h = W_h^V e_{m_2}, \quad (3)$$

Where $m_1, m_2 \in \{e, p, s\}$ represents the modality used. In each subspace, scaled dot product attention operations were performed on these projections. For subspace $h$, the attention operation expression is as follows:

$$Att_h(Q_h, K_h, V_h) = Softmax\left(\frac{Q_h K_h}{\sqrt{d_k}}\right) V_h \quad (4)$$

Where $Att_h(\cdot)$ refers to attentional operations in subspace h, and $d_k$ is the characteristic dimension. All H attention outputs are connected in series and passed through a linear layer to obtain the final output of the multi-head attention (MHA) module.

To achieve an effective fusion of different modalities, we input the features of any two modalities into MHA and obtained feature weights that contain common information between these two modalities. Subsequently, we perform a mean operation to obtain the final feature weights $\omega_{m1m2}$ for these two modalities, as follows:

$$\omega_{m1m2} = \mu(MHA(f_{m1}, f_{m2}, f_{m2})) \quad (5)$$

Where $f_{m1}, f_{m2} \in \{f_e, f_p, f_s\}$ represents the features of any two modalities. The feature weights of any two modes are calculated to represent: $\omega_{es}, \omega_{ps}, \omega_{se}, \omega_{pe}, \omega_{sp}, \omega_{ep}$. Finally, these weights information are stacked to achieve an effective fusion of the three modalities of eye movement, PPG, and video semantic information, obtained feature weights $\omega_{eps}$ containing important information from three modalities:

$$\omega_{eps} = [\omega_{es}, \omega_{ps}, \omega_{se}, \omega_{pe}, \omega_{sp}, \omega_{ep}] \quad (6)$$

Finally, the fused multimodal features are fed into the emotion classifier for emotion prediction. The classifier is composed of two fully connected layers, and the softmax activation function is used in the second fully connected layer. The mathematical expression of emotion prediction is as follows:

$$\hat{y} = Softmax\left(FC_{\theta_2}\left(FC_{\theta_1}[\mu + \sigma]\right)\right) \quad (7)$$

Where $\hat{y}$ represents the final emotion prediction result, $FC_{\theta_1}$ and $FC_{\theta_2}$ represent fully connected layers with parameters $\theta_1$ and $\theta_2$, respectively, $\mu$ and $\sigma$ are the average and standard deviation calculated from the output $\omega_{eps}$ of the fusion module, and + represents the concatenation operation.

## 4 EXPERIMENTS

### 4.1 Dataset and Data Processing

To verify the effectiveness of our proposed method, we conducted experiments on the Video Learning Multimodal Emotion Dataset (VLMED) [3, 37], which contained the subjects' eye movement, PPG, facial expression, EDA data and the instructional videos watched by the subjects. The data was collected while the subjects watched instructional videos. This dataset simulates MOOC learning scenarios during the collection process, using 5 carefully selected instructional videos to induce different types of emotions: interest, boredom, happiness, confusion, and distraction. The experiment collected data from 68 subjects, each of whom watched 5 videos in sequence, including 4 shorter (about 2-3 minutes) and 1 longer (about 10 minutes) instructional video.

In this study, we mainly used eye movement, PPG data, and instructional videos from this dataset. Extract data with a time window of 1 second, and process and extract features from eye movement and PPG data using the same methods as in papers [3] and [37], respectively. We also extracted semantic information from instructional videos to expand the data set. The acquisition method of video semantic information and its feature extraction are introduced in Section 3.2.

During the experiment, we observed that the model performed very poorly in recognizing the emotion of Interest category, and the same problem was encountered in the work of literature 1 [3] and literature 2 [24]. We speculate that it may be caused by unbalanced samples in the data set. To solve this problem, we adopted the ADASYN sampling approach [34] to enhance the data. ADASYN is a data resampling-based method that synthesizes small sample categories in the feature space to generate high-quality new samples, thereby balancing the distribution of samples in different categories. The number of samples before and after adaptive synthesis sampling is shown in Table 4.

Table 1. Sample size before and after using ADASYN

| Data | Interest | Boredom | Happiness | Confusion |
| --- | --- | --- | --- | --- |
| Raw data | 1451 | 2723 | 1761 | 2275 |
| ADASYN | 2848 | 2807 | 2723 | 2605 |

## 4.2 Experimental Setting

In this study, we used NVIDIA GeForce RTX 3070 GPU as the computing platform and constructed and trained the entire model using the TensorFlow framework. We divided the dataset into training and testing sets according to the ratio of 8:2 and trained the model using 5-fold cross-validation on the training set. At the same time, we evaluated the performance of the model using the testing set. During the experiment, we attempted different parameter combinations to determine the optimal parameter configuration as the final parameters of the model. The final network and training parameters of the model are set as follows:

**Network Parameters**：In the data processing and feature extraction module, we used a Conv1D and a LSTM network to encode eye movement and PPG features. Conv1D includes 16 filters of size 1 and uses ReLU as the activation function; LSTM contains 64 hidden units. Encode semantic features using a LSTM with 64 hidden units. In the cross-attention fusion module, num\_heads=8, key\_dim=128, and value\_dim=64 in multi-head attention. The emotion classifier consists of two fully connected layers, the first consisting of 64 units, while the second consists of 4 units and uses the softmax activation function to achieve the classification of 4 emotions. In addition, we have introduced L2 regularization (l2=0.001) at various levels of the network to reduce model complexity, prevent overfitting, and enhance the model's generalization ability.

**Training Parameters**：When training, we use sparse categorical cross-entropy as the loss function, Adam as the optimizer, and set the random seed to 7 to ensure the repeatability of the experimental results. Set batch size = 32, epoch = 500, and learning rate= 1e-3. To avoid overfitting, we set the learning rate decay and early stop criteria for model training. If the model does not show improvement for 5 consecutive epochs, the learning rate is attenuated to the original 0.1. When the model does not show better performance for 10 consecutive epochs, we determine that the model is overfitted and terminate the training.

To evaluate the effectiveness of the model, we comprehensively tested its performance using 5-fold cross-validation. We calculated the average accuracy ($Avg_{acc}$), average recall ($Avg_{recall}$), and average F1 score ($Avg_{f1}$) as evaluation metrics.

## 4.3 Results and analysis

Figure 3 shows the confusion matrix for each fold of our model under 5-fold cross-validation. It can be observed that the difference between the results of each fold is not large, which indicates that our proposed model shows effective and stable performance in MOOC learning scenarios. However, we found that the model had relatively low accuracy in identifying the two categories of Interest and Confusion. Specifically, the Interest category is easily misclassified as either Happiness or Confusion, which may be due to both Interest and Happiness representing positive emotions, so they are easy to confuse when classifying emotions. Similarly, Confusion and Boredom are both negative emotions, resulting in some Confusion samples being incorrectly classified as Boredom. In addition, compared with Happiness and Boredom, Interest and Confusion are neutral emotions with low emotional intensity, and their emotion scores are similar. Therefore, some samples of Interest and Confusion have similar feature distributions on the two physiological signals of eye movement and PPG, which is difficult to distinguish effectively. The above analysis is confirmed in the visualization results of feature distribution in Figure 5.

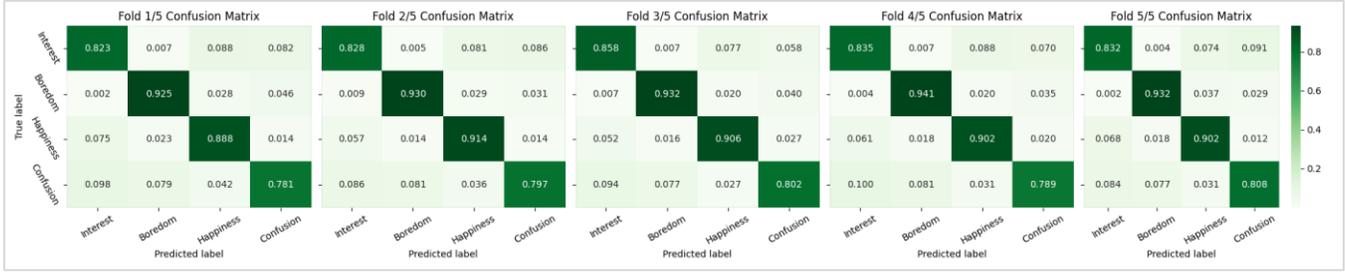

Figure 3. The confusion matrix of each fold of the model under 5-fold cross-validation

### 4.3.1 Effectiveness of adaptive synthetic sampling.
To overcome the potential impact of imbalanced data distribution, we adopted the Adaptive Synthesis (ADASYN) sampling method for data augmentation, and its effectiveness was verified through experiments. The experimental results are shown in Figure 4, where (a) and (b) are the ROC curves before and after using ADASYN, respectively. It can be found that without ADASYN processing, the model performs poorly in recognizing the emotion of Interest category. After ADASYN processing, the model has significantly improved its recognition of various emotions. This method effectively alleviates the problem caused by the unbalanced data distribution and improves the model's overall performance. It should be emphasized that the enhanced data were used in other experiments in this paper.

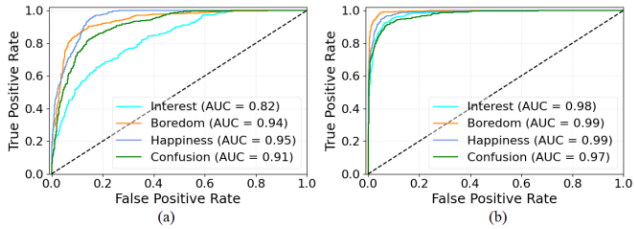

Figure 4. The ROC curves with (a) and without (b) ADASYN

### 4.3.2 Compare with other models.
To better demonstrate the effectiveness of our model, using the data collected in this paper, we reproduce six baseline classifiers and compare them with the methods proposed in this paper, including the traditional machine learning method K nearest neighbor (KNN) [6], deep learning methods of LSTM [1], CNN-LSTM[5], as well as Transformer [31], CNN-LSTM-MHA-TCN (CLA-TCN) [37], and Cascade Multi-Head Attention(CMHA)[39] using attention mechanisms. The results are shown in Table 2. The effect obtained by using the deep learning method is significantly better than that obtained by the machine learning method, indicating that the deep learning method can extract deeper features. When the attention mechanism is used, the effect is further improved, and our method achieves the best performance, indicating that our model can effectively learn important information between different modalities and more efficiently fuse information from different modalities.

Table 2. Results compared with other models

| Model | Acc±std(%) | Recall(%) | F1 |
|---|---|---|---|
| KNN[6] | 59.56±2.7 | 59.67 | 0.58 |
| LSTM[1] | 67.59±2.0 | 67.89 | 0.67 |
| CNN-LSTM[5] | 73.23±1.7 | 73.14 | 0.73 |
| Transformer[31] | 78.72±1.8 | 78.96 | 0.78 |
| CLA-TCN[37] | 82.52±2.1 | 81.03 | 0.81 |
| CMHA[39] | 83.97±1.2 | 84.01 | 0.84 |
| Ours | 86.69±0.7 | 86.62 | 0.87 |

### 4.3.3 Comparison with different semantic information.
Unlike most studies that transcribe audio into text as semantic information, we use generated video descriptions as semantic information. To demonstrate the effectiveness of the proposed method, we conducted experiments using caption semantic (Audio transcription into text subtitles as semantic information) and semantic information generated by BiliGPT (https://bibigpt.co, First transcribe the audio into text and then further summarize it as semantic information). Table 3 shows the experimental results. We can see that the emotion recognition effect is improved after using subtitle semantics, but it is not as good as the summary subtitle with reduced redundant information after the summary. When description semantics is used, better results are obtained, because the learners' emotional production in the learning process is affected by the visual content stimulation, and the video description contains this information. Therefore, using video description as semantic information can get better results, which also reflects the innovation of the method in this paper.

Table 3. The results of using different semantic information

| Semantic Type | Acc ± std(%) | Recall(%) | F1 |
|---|---|---|---|
| Without Semantic | 63.57±1.9 | 63.58 | 0.64 |
| Caption Semantic | 76.12±1.6 | 76.17 | 0.76 |
| Caption Summary Semantic | 78.44±1.5 | 78.54 | 0.79 |
| Description Semantic | 86.69±0.7 | 86.62 | 0.86 |

## 4.4 Ablation Studies

### 4.4.1 Effectiveness of multimodal fusion.
We conducted experiments using unimodal, bimodal, and trimodal, and the results are shown in Table 4. We observed that emotion recognition improved significantly when multimodal data was used. This result shows that integrating multi-modal data helps to capture learners' emotional states more comprehensively, thus achieving higher performance affective perception. In addition, comparing experiments I and II in Table 4, we found that the use of eye movement could obtain better results than PPG signals, indicating that there is a strong correlation between learners' emotions and eye movement signals, possibly because in MOOC learning scenarios, learners mainly watch instructional videos through vision. This result also suggests that learners' emotions can be affected by visual stimuli.

*4.4.2 Effectiveness of Video Semantics.* As shown in Table 4, the experimental effect has been significantly improved after incorporating video semantic information, whether it is bimodal or trimodal. Further comparing experiments IV and V in Table 4, we found that using eye movement signals and video semantic information for emotion recognition is more effective than using PPG signals and video semantic information. This indicates a stronger correlation between eye movement signals and semantic information. In MOOC learning scenarios, learners acquire knowledge by watching instructional videos, so eye movement signals are naturally directly affected by the instructional videos. This also confirms the importance of integrating video semantic information into MOOC learning emotion recognition tasks.

*4.4.3 Effectiveness of cross-attention.* To verify the effectiveness of our proposed cross-attention fusion method, we also conducted experiments by directly concatenating features without using cross-attention fusion. The experimental results show that when cross-attention is used, the accuracy of emotion recognition is significantly improved (See experiments VI and VII in Table 4). This indicates that our model can effectively capture the correlation and complementarity between different modalities, thereby improving the performance of emotion recognition. In our method, the data from three modalities is combined pairwise and fed into MHA, so that each modality can learn information related to the other two modalities. Then, the learned features are further fused to obtain features containing important information about the three modalities. Finally, the fused features are used for emotion recognition. By this method, we achieved the effective fusion of multimodal data, which can make full use of the effective information of each modality and improve the accuracy of emotion recognition.

## 4.5 Effects on public dataset

To demonstrate the generalization ability of our method, we conducted experiments on the publicly available dataset MAHNOB-HCI [25]. This dataset is a multimodal database that synchronously records the data of 27 subjects' EEG, eye movements, facial video, audio signals, and peripheral physiological signals while they watched 20 emotional videos. We used the EEG and eye movement signals, along with extracted semantic information from the videos in this dataset, for experimentation. We compared the results in terms of arousal (including Calm, Medium arousal, Excited/Activated) and valence (including Unpleasant, Neutral valence, and Pleasant) dimensions with the baseline. The experimental results are shown in Table 5. We can see that compared to using single-modal EEG and eye-tracking data, the performance significantly improves when adding video semantic information. The best results are achieved when using all three modalities simultaneously. This experiment further demonstrates the positive impact of incorporating video semantic information on emotion recognition. It also validates the strong generalization capability of our method, making it suitable for emotion recognition tasks induced by stimulus materials.

Table 5. Experimental results using EEG, Eye Movement (EM), and Video Semantic Information (VSI) in the MAHNOB-HCI

| Model | Modality | Acc (%) | | F1-score | |
|---|---|---|---|---|---|
| | | arousal | valence | arousal | valence |
| Baseline | EEG | 52.4 | 57.0 | 0.42 | 0.56 |
| | EM | 63.5 | 68.8 | 0.60 | 0.68 |
| | EEG & EM | 67.7 | 76.1 | 0.62 | 0.74 |
| Ours | EEG | 53.2 | 55.9 | 0.49 | 0.53 |
| | EM | 62.9 | 68.4 | 0.57 | 0.63 |
| | EEG & VSI | 67.1 | 72.4 | 0.61 | 0.64 |
| | EM & VSI | 68.8 | 73.9 | 0.67 | 0.71 |
| | EEG & EM & VSI | **82.3** | **82.8** | **0.80** | **0.79** |

## 4.6 Visualization

To demonstrate the effectiveness of our method more clearly, we visualized the feature distributions learned by the classifier in the second-to-last layer of our model using t-SNE [17] in three different settings. As shown in Figure 5 (a), it is difficult for the model to effectively distinguish different categories of emotions using only eye movement and PPG data. As shown in Figure 5 (b), with the addition of video semantic information, it can be observed that the feature distribution distinguishes different emotion categories becomes more clear, which further verifies that fusing video semantic information has a positive effect on improving emotion recognition performance. When the cross-attention mechanism is further applied, the feature distribution becomes more obvious (as shown in Figure 5 (c)), indicating that the cross-attention mechanism can effectively learn information between different modalities and improve emotion recognition performance. In addition, in Figure 5, we can also find that it is difficult to distinguish the emotional categories of Interest and Confusion, which explains why the recognition accuracy of Interest and Confusion is low (as shown in Figure 3 and Figure 4 (a)).

Table 4. The experimental results of using eye movement (EM), PPG, and video semantic information (VSI) data, as well as whether the cross-attention mechanism (CA) is used

| Experiment number | Modality | | | CA | Acc ± std (%) | Recall (%) | F1-score | Params(MB) |
|---|---|---|---|---|---|---|---|---|
| | EM | PPG | VSI | | | | | |
| I | √ | - | - | - | 60.15 ± 3.8 | 60.00 | 0.60 | 0.696 |
| II | - | √ | - | - | 44.62 ± 0.6 | 40.79 | 0.39 | 0.686 |
| III | √ | √ | - | √ | 63.57 ± 1.9 | 63.54 | 0.61 | 0.764 |
| IV | - | √ | √ | √ | 69.88 ± 0.6 | 69.87 | 0.69 | 0.763 |
| V | √ | - | √ | √ | 72.77 ± 1.5 | 72.79 | 0.73 | 0.764 |
| VI | √ | √ | √ | - | 72.10 ± 1.7 | 72.23 | 0.72 | 0.646 |
| VII | √ | √ | √ | √ | **86.69 ± 0.7** | **86.62** | **0.87** | **1.420** |

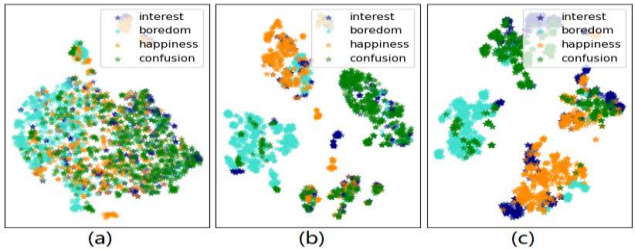

Figure 5. Visualization results of feature distribution in three settings. In Figure (a), only eye movement and PPG data are used. In Figure (b), video semantic information is further integrated on the basis of (a), but no cross-attention mechanism is adopted; In Figure (c), on the basis of (b), the cross-attention mechanism is further introduced

## 5 Conclusions and Future Works

In this work, we propose a multimodal emotion recognition method that integrates video semantic information and physiological signals, aiming at the particularity of the MOOC learning scenario. This is the first attempt to apply semantic information from instructional videos to emotion recognition tasks in MOOC learning. We use a method of generating video descriptions to extract high-level semantic information from educational videos, thereby expanding the dataset. Experimental results indicate that incorporating video semantic information has a significantly positive impact on emotion recognition. We use cross-attention to capture semantic correlations between different sequences and have designed a multimodal fusion method based on cross-attention. This method successfully fuses video semantic information with physiological signals, achieving an accuracy improvement of over 14%. Additionally, we adopted adaptive synthetic sampling for data augmentation, effectively eliminating the impact of data distribution imbalance. To validate the generalization ability of our approach, we further conducted experiments on the publicly available HCI dataset. The results indicate that our method can significantly improve the performance of emotion recognition. Overall, through extensive experimentation, we have demonstrated the effectiveness and feasibility of the proposed method, providing new perspectives and effective approaches for emotion recognition studies induced by stimuli materials.

In the current study, we used the global semantic information of the instructional video for analysis. In future work, we will try to extract more fine-grained video semantic information to conduct experiments. In addition, we will also explore more effective multi-modal fusion strategies to fully utilize information from different modalities to achieve higher emotion recognition performance.

## ACKNOWLEDGMENTS

We sincerely appreciate all the editors and reviewers for their insightful comments and constructive suggestions. This research work was supported by National Natural Science Foundation of China under Grant No.62267001 and No.62307009.

## Data Availability Statemen

Due to privacy, copyright, commercial confidentiality, or legal restrictions, the dataset used in this article is not publicly available. If you have specific research or analysis needs for this data set, you can contact the maintainer or relevant person in charge of the data set through the following link: https://github.com/zhou9794/video-learning-multimodal-emotion-dataset. Typically, you may need to sign a confidentiality agreement or a data use agreement to gain access to the dataset.